# Atomic transfers between implanted bioceramics and tissues in orthopaedics surgery

## J.L. IRIGARAY, E. CHASSOT, G. GUIBERT, E. JALLOT

Physics Corpuscular Laboratory, IN2P3 – CNRS, Blaise Pascal University of Clermont-Ferrand, F – 63177 Aubière Cedex, France

## Introduction

We study transfers of ions and debris from bioceramics implanted in bone sites. A contamination of surrounding tissues may play a major role in aseptic loosening of the implant. For these reasons, bioceramics require studies of biocompatibility and biofunctionality[1]. So, in addition to *in vitro* studies of bioceramics, it is essential to implant them *in vivo* to know body reactions.

We measured the concentration of mineral elements at different time intervals after implantation over a whole cross-section. We found a discontinuity of the mineral elements (Ca, P, Sr, Zn, Fe) at the interface between the implant and the receiver. The osseous attack is not global but, on the contrary, centripetal. Moreover, the fit of the concentration time course indicates that the kinetics of ossification is different for each atomic element and characterizes a distinct biological phenomenon.

## Materials

Several kinds of materials have been studied in our research group during last years. Some ones are natural, the others are synthetical or composites. However, all of them are implanted in bone sites to study their application in orthopaedic surgery. First of all, the biological coral is a natural material, treated chemically for implantation "in vivo" sites. It is *porites lutéa*, with 5% of porosity, and crystallized in orthorhombic form. It is sterilized either by thermal effect either by photonic irradiation. The samples are cylinders of about 5 mm diameter and 10 mm long.

The second kind is hydroxyapatites.
Pure material, $Ca_{10}(PO_4)_6(OH)_2$, is noted HA.
Composites are made by including two metallic elements: doped with manganese is noted (HA, Mn), and doped with zinc is noted (HA, Zn).
We have also studied the HA composite made with 75% HA and 25% of β-tricalcium phosphate.
For all them, the porosity is 50% and pores diameter spreads from 336 to 450 µm.
The last kind is biological glasses. They contain different oxides in appropriate percentages: $SiO_2$, $P_2O_5$, CaO, MgO, $Na_2O$, $K_2O$ and $Al_2O_3$.
They are not manufactured as cylinders but they are used as powders for coating metallic cylinders Ti6Al4V by plasma spraying method. The thickness of the coatings is not regular and extends between 30 to 100 µm.
By modelling oxides percentages, we have obtained and studied bioactive and inert glasses.

## Methods

Nods are implanted into lateral femoral epiphysis of sheeps by manual pressure. Animals are sacrificed at some time intervals during one year. After extraction of samples, analyses are made by several means: Scanning Electron Microscopy (SEM), Scanning Transmission Electron Microscopy

(STEM), Energy Dispersive X-ray Spectroscopy (EDXS), Particle Induced X-ray Emission (PIXE) and Rutherford Back Scattering (RBS).

Cross-sections are observed using a scanning transmission electron microprobe operating at a voltage of 100 kV.

Elemental profiles at interface are performed using energy dispersive X-ray spectroscopy. The detector was a Si(Li) diode equipped with an ultra thin window. After implantation of the bioceramic, STEM image and the concentrations profiles across the bioceramic and bone interface permit to distinguish different zones at the implant periphery under the micrometer scale.

However, when one is interested by quantitative and very sensitive measurements, it is valuable to develop nuclear instruments and methods. Proton induced X-ray emission is applied to cartography selected elements near the interfaces.

These methods have been published elsewhere[2,3,4]. A scanning of samples has been made with proton beams to obtain cartographies of interfaces between implants and tissues. In this case, it is necessary to take account carefully of the continuous background in PIXE method[5].

**Results**

After extraction from the thighbone, samples show that coral is ossified after a delay of 5 months (see figures 1). However, the use of methyldiphosphonate molecules labelled with radioactive 99mTc specifies that the biofunctionality is reached only after a delay of 8 months.

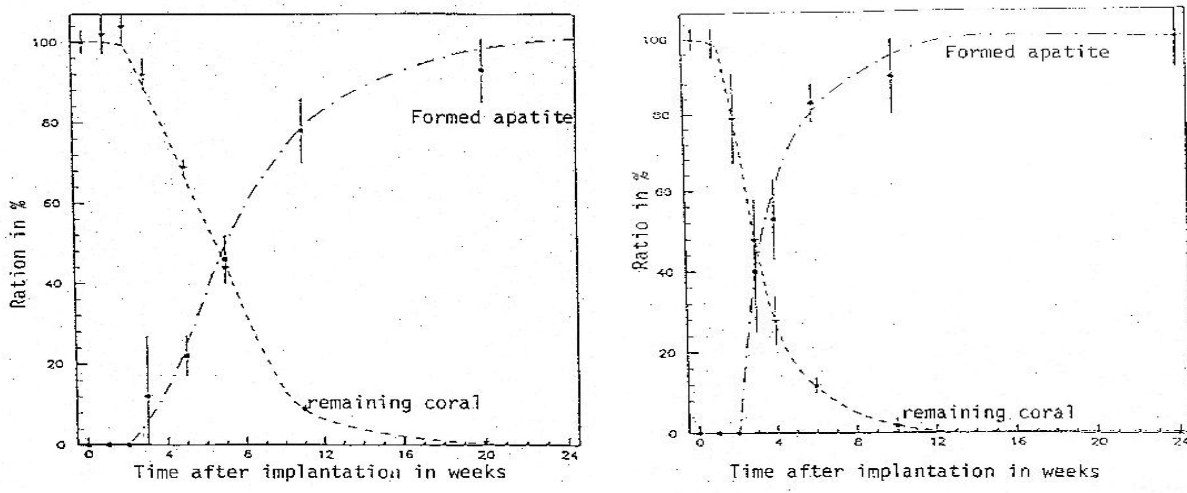

Figure 1: Kinetics of coral resorption and apatite formation at ovines (left) and procines (right)

We can see also that the kinetics of coral resorption and apatite formation is not the same for ovines and porcines.

In figure 2, the diagrams show the implantation of the sample in bone, the proton beam scanning and an example of cartography.

We have observed that, after 48 weeks of implantation of the sample in bone, the remaining percentage of HA in the implanted bioceramics is: 19% for pure HA, 20% for HA doped with Mn, 7% for HA doped with β-tricalcium phosphate and only 1% for HA doped with Zn. Mn has no effect but Zn accelerates the resorption kinetic.

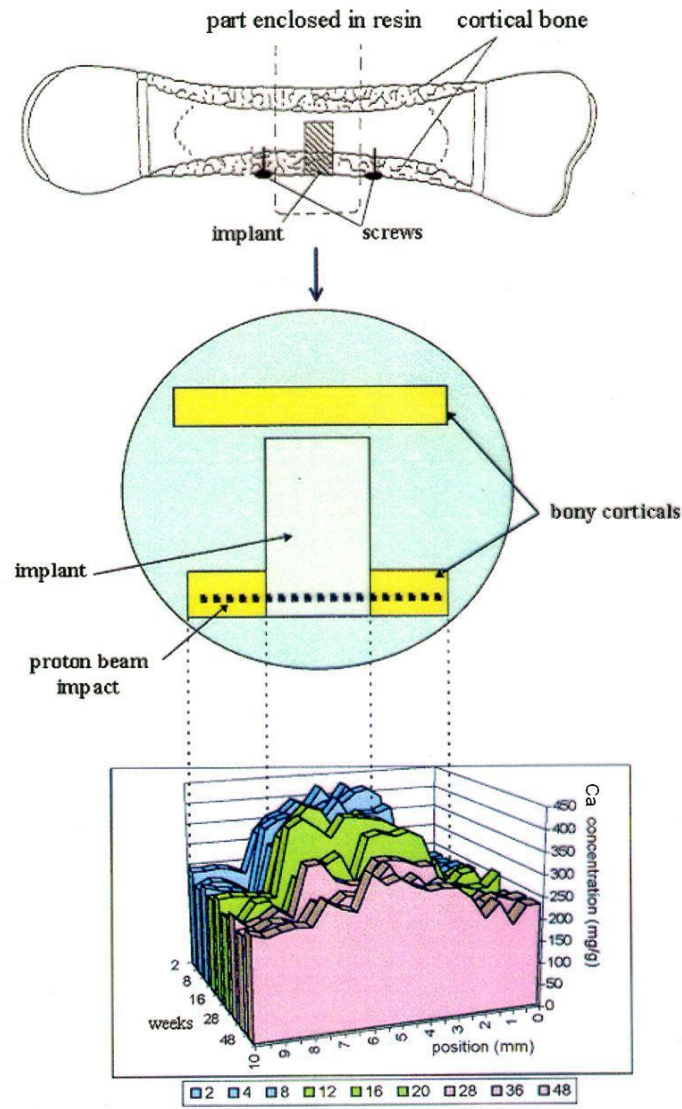

Figure 2: Calcium distribution in extracted biopsies of bone-HA implant-bone versus time after implantation.

In the case of biological glasses, yet after 3 months of implantation, they are transformed into silicon gel with incorporation of proteins and trace elements such as Zn and Sr (figure 3).

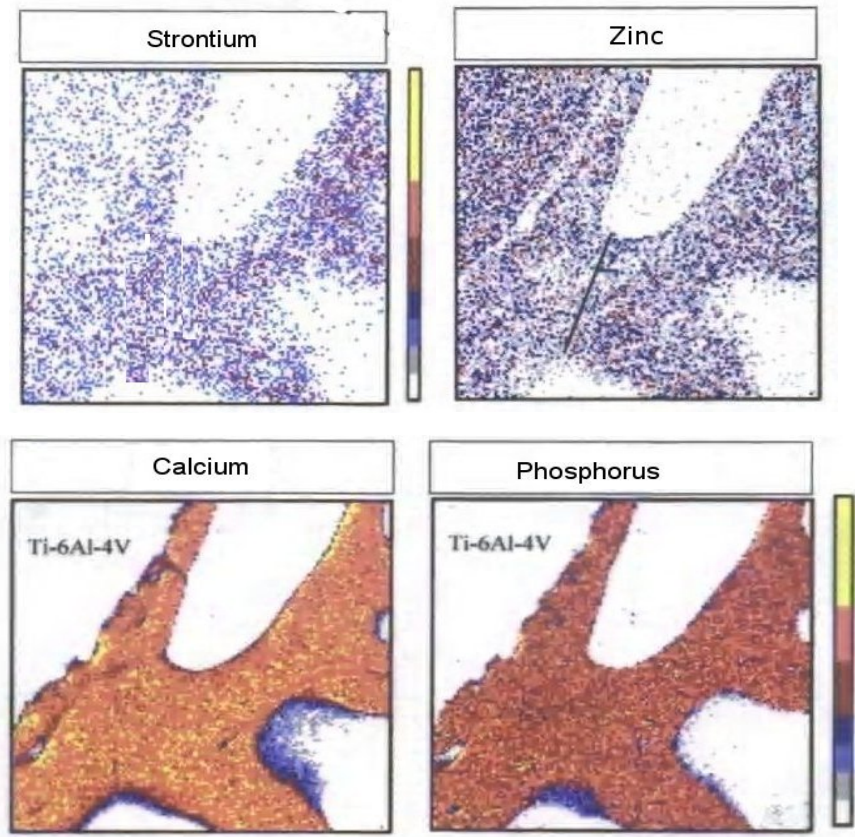

Figure 3: PIXE cartographies on a bioactive glass after 6 months of implantation in previously selected region by SEM.

This gel disappears gradually and is replaced by neoformed bone. Formation of interface induces migration of glass particles through the lacuna network of surrounding bones and then produces an effect on osteoblasts proliferation[6]. A bioactive glass coating increases the bone contact perimeter and the neoformed bone becomes more quickly mature. Metallic prosthesis is better integrated in bone tissues. We can expect that this improvement involves a decrease of micro-motions which are source of contaminations by metal elements.

We observe that bioactive glasses can be effective barrier against corrosion of alloy as long as they remain in place. After a few months of implantation, the titanium contamination remains localized within the first tens of micrometers of surrounding bone.

**Conclusions**

Several ceramics can be adapted for orthopaedic surgery. Coral is rapidly transformed into neoformed bone for bone repair.

Hydroxyapatite and its composites are convenient for coating of metallic implants to obtain a good adherence. Large theoretical studies are carried on to adapt the response of the crystalline structure versus the applications[7]. The physico-chemical and biological properties of apatites can be modified by ions substitutions in these apatites.

The most interesting possibilities are provided by nanocrystalline apatites. They are characterized by the existence on the crystal surface of a hydrated layer of loosely bound mineral ions which can be easily exchanged in solution.

Biological glasses improve osseointegration of prosthesis when they are bioactive or are employed like cement when they are bioinert.